\documentclass{article}

\usepackage{PRIMEarxiv}

\usepackage[utf8]{inputenc} 
\usepackage[T1]{fontenc}    
\usepackage{hyperref}       
\usepackage{url}            
\usepackage{booktabs}       
\usepackage{amsfonts}       
\usepackage{nicefrac}       
\usepackage{microtype}      
\usepackage{lipsum}
\usepackage{fancyhdr}       
\usepackage{graphicx}       
\graphicspath{{media/}}     
\usepackage{comment}
\usepackage{tabularx}

\newcommand{\subhead}[1]{\vspace{0.5em}\noindent\textbf{#1.}}

\newcommand{\tabfontsize}{}

\title{Synthetic
Audio Forensics Evaluation (SAFE) Challenge}

\begin{document}

\author{
  Kirill Trapeznikov, Paul Cummer, Pranay Pherwani, Jai Aslam \\
  STR \\
  Woburn, MA \\
  \texttt{\{kirill.trapeznikov, paul.cummer, pranay.pherwani, jai.aslam\}@str.us} \\
  \And
  Michael S. Davinroy, Peter Bautista, Laura Cassani \\
  Aptima, Inc. \\
  Woburn, MA \\
  \texttt{\{mdavinroy, pbautista, lcassani\}@aptima.com} \\
  \And
  Matthew Stamm \\
  Drexel University \\
  Philadelphia, PA \\
  \texttt{mstamm@drexel.edu} \\
  \And
  Jill Crisman \\
  ULRI Digital Safety Research Institute \\
  Northbrook, IL \\
  \texttt{jill.crisman@ul.org} \\
}



%

\maketitle

\begin{abstract}
The increasing realism of synthetic speech generated by advanced text-to-speech (TTS) models, coupled with post-processing and laundering techniques, presents a significant challenge for audio forensic detection. In this paper, we introduce the SAFE (Synthetic Audio Forensics Evaluation) Challenge, a fully blind evaluation framework designed to benchmark detection models across progressively harder scenarios: raw synthetic speech, processed audio (e.g., compression, resampling), and laundered audio intended to evade forensic analysis. The SAFE challenge consisted of a total of 90 hours of audio and 21,000 audio samples split across 21 different real sources and 17 different TTS models and 3 tasks. We present the challenge, evaluation design and tasks, dataset details, and initial insights into the strengths and limitations of current approaches, offering a foundation for advancing synthetic audio detection research. More information is available at \href{https://stresearch.github.io/SAFE/}{https://stresearch.github.io/SAFE/}.
\end{abstract}

\begin{figure}[htb]
\centering
    \fbox{\includegraphics[width=.95 \textwidth]
    {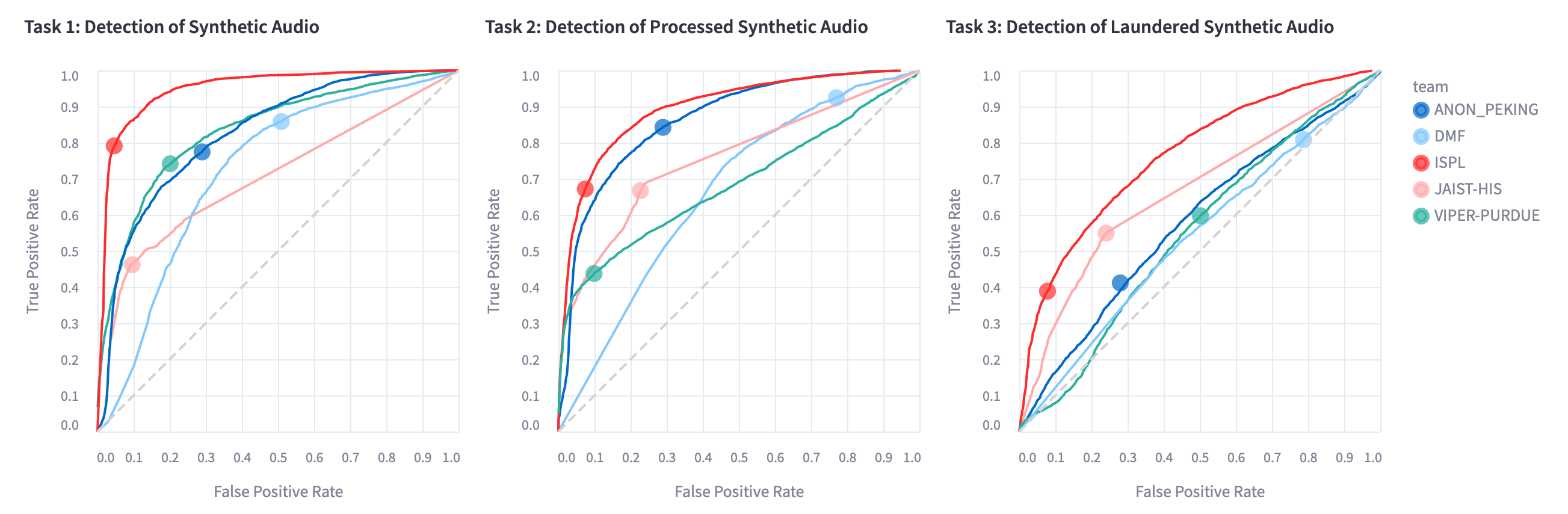}}
    \caption{Synthetic
Audio Forensics Evaluation Challenge Round 1 Results. 
Performance (circle markers) from top five teams and the detection vs. false alarm curves are shown for three tasks of increasing difficulty: detection of (1) synthetic voice audio, (2) synthetic voice audio post-processed with various compression and resampling and (3) laundered to evade detection.} 
    \label{fig:teaser}
\end{figure}

\section{Introduction}
Recent advances in synthetic audio generation that are driven by increasingly sophisticated text-to-speech (TTS) models and speech manipulation techniques pose significant challenges to the authenticity and trustworthiness of audio content. As synthetic speech becomes more realistic and widely accessible, malicious actors are increasingly able to create convincing forgeries that can undermine public trust, enable fraud and impersonation, and threaten security-sensitive applications. 
As a result, forensic detection methods must evolve to keep pace with the growing threat of synthetic audio, particularly in scenarios involving post-processing, compression, or intentional laundering designed to evade forensic analysis. 

To address these growing threats, we created the SAFE (Synthetic Audio Forensic Evaluation) Challenge. SAFE emphasizes critical research areas that include robustness across diverse data sources (including both generated and real audio), resilience to emerging laundering techniques, and computational efficiency for practical deployment considerations. SAFE aims to provide a rigorous, blind evaluation framework that targets three key areas of forensic analysis: detection of 1) raw synthetic speech, 2) compressed and resampled synthetic audio, and 3) audio subjected to laundering attacks intended to obfuscate synthetic origins. By benchmarking performance across a diverse, balanced set of real and synthetic sources under controlled computational conditions, SAFE aimed to evaluate not only detection accuracy, but also generalization across unseen sources and resilience to realistic adversarial conditions. 
This competition makes the following key contributions: 
\begin{itemize}
    \item \textbf{Blind Evaluation Protocol}: We designed and deployed a fully blind evaluation framework where participants had no access to evaluation data, ensuring an unbiased assessment of model generalization to unseen audio sources and manipulations. During Round 1 (70 days) of the competition, we received more than 700 submissions from 12 teams.
    \item \textbf{Benchmarking Against Processing and Laundering}: We introduced benchmark tasks specifically focused on detection of synthetic audio subjected to common post-processing (e.g., compression, resampling) and laundering intended to bypass detection. 
    \item \textbf{Source-Balanced Dataset Design}: We constructed a balanced evaluation set to fairly assess model performance across a diverse range of human and machine-generated audio comprising 13  TTS generative models and 21 real sources for a total of 90 hours and 22,700 audio samples. 
    \item \textbf{Performance Analysis}: We provided fine-grained anonymized performance  across sources and augmentation methods, offering deeper insights into model strengths and weaknesses.
\end{itemize}
\section{Background}

Rapid developments in generative AI have led to the creation of 
AI systems capable of creating realistic synthetic speech~\cite{barrington2025people}.  In response to this, a number of forensic systems have been developed to detect synthetic audio, and several datasets and challenges have been designed to further research in this area.


\subhead{Synthetic Audio Datasets}
High-quality datasets containing both real and synthetic audio are essential for advancing research on synthetic audio detection and for benchmarking the performance of detection systems. To keep pace with the rapidly evolving landscape of generative audio technology, several synthetic audio datasets have been developed. Among the earliest are those created for the ASVspoof challenges~\cite{todisco2019asvspoof,todisco2019asvspoof}, 
which focus on spoofed audio for automatic speaker verification. Salvi et al.~\cite{salvi_timit-tts_2023} introduced the TIMIT-TTS dataset to provide realistic synthetic speech tracks aligned with deepfake videos for audio and multimodal forensic analysis, using a pipeline that integrates text-to-speech synthesis with dynamic time warping. More recently, Bhagtani et al.~\cite{bhagtani2025diffssd} released the DiffSSD dataset to fill a critical gap by offering synthetic audio generated using diffusion-based methods—an increasingly prevalent class of generative models.

\subhead{Synthetic Audio Identification Competitions} 
A range of challenges and competitions have emerged in response to the growing prevalence of AI-generated or ``deepfake" audio. The ASVspoof challenges~\cite{todisco2019asvspoof,yamagishi2021asvspoof,liu_asvspoof_2023} were among the earliest of these challenges to arise. These challenges primarily focused on speaker verification, i.e. determining whether speech was produced by a known speaker or by using a synthetic audio system to spoof that speaker, as opposed to detecting synthetic audio from an unknown source. 
%
Several synthetic audio detectors ~\cite{corvey2024semantic} were also developed during the DARPA Semantic Forensics (SemaFor) program, however, evaluations within this program were limited to program performers and were not accessible to the wider research community. Finally, in 2022, the IEEE Signal Processing Cup hosted a competition that tasked participants with synthetic audio attribution~\cite{salvi2023synthetic}, but its primary audience was undergraduate students, limiting its impact on the broader research field. 
In contrast, the SAFE competition was open to the full research community and pursued two main objectives: (1) to catalyze the development of robust synthetic audio detection systems capable of generalizing across unknown speakers, and (2) to identify persistent challenges and performance gaps in current detection methodologies.




\section{Competition Overview}
The SAFE Challenge aimed to mobilize the research community to advance the state of the art in audio voice forensics and drive innovation in detecting synthetic and manipulated audio artifacts. The challenge focused on several critical aspects including generalizability across diverse audio sources and newly emerging synthesis models, robustness to benign post-processing and targeted laundering methods. The competition was intentionally designed to mimic real-world setting. Participants did not have access to the dataset and were not aware of the specific sources or generative models used in data creation. We provided only minimal feedback in the form of top-level accuracy across a subset of anonymized sources in the public split.

Calibrating the difficulty of the competition was guided by several key principles. First, the tasks were intentionally structured to increase in difficulty, ensuring a progression from basic detection to more challenging, real-world adversarial scenarios. To achieve this difficulty gradient, we chose task objectives and underlying datasets that represent semantically different, and increasingly more complicated and adversarial, problem spaces. The first task consisted of raw synthetically generated audio; the second task consisted of synthetically generated audio compressed in differing ways, adding simulated complexity of real-world conditions; and the third task consisted of an adversarial approach that attempted to fool participant detectors by adding various background noise, which we refer to as laundering attacks. Second, we designed the difficulty of the first task to be balanced on two baseline audio detectors developed under the DARPA SemaFor program~\cite{corvey2024semantic}.  
Setting the balanced accuracy of both models to $.83$ ensured that the first task was challenging enough to allow for meaningful improvements in detection model design but not so difficult that participants could not make progress given the limited feedback.

\subsection{Tasks}
The SAFE Challenge was composed of three related tasks of increasing difficulty:

{\bf Task 1. Detection of synthetic audio} focused purely on the detection of generated voice audio from popular TTS models. No modifications or post-processing were applied to the generated audio or the real sources. Information about the 21 real audio sources and 13 synthetic models are shown in Table \ref{tab:real} and Table~\ref{tab:generated} respectively.


{\bf Task 2. Detection of processed synthetic audio} tested how detection performance is affected by common forms of post-processing, such as audio compression codecs and resampling. The post-processing only applied to generated audio, while the real data from Task 1 remained the same. Table~\ref{tab:augmentation_detection} lists the 19 different post-processing 
methods used.


{\bf Task 3. Detection of laundered synthetic audio} tested detection performance on laundered audio. In this task, 
the generated audio files were laundered 
to purposefully evade detection, while the real dataset remained unaltered.
Four different laundering methods shown in Table~\ref{tab:laundering_descr} were applied.

\begin{table}[t]
\centering

\begin{tabular}{lclc}

\toprule
\textbf{Source} & \textbf{SR(kHz)} & \textbf{Description} & \textbf{Public Split} \\
\midrule
Mandarin Podcast 1       & 48        & Podcast in Mandarin          & yes \\
Arabic Speech Corpus \cite{halabi2016arabic}     & 48        & Arabic speech                & yes \\
English Podcast           & 44.1      & Podcast in English           & yes \\
Conference                & 48        & Conference speakers          & yes \\
Fleurs German \cite{conneau2023fleurs}             & 16        & German research data         & yes \\
High Quality Podcasts     & 44.1, 48  & High quality podcasts        & yes \\
Japanese Shortwave        & 12        & Japanese shortwave radio     & yes \\
VSP Documentary           & 44.1      & Documentary audio            & yes \\
VSP Phone Call            & 44.1      & Phone call audio             & yes \\
VSP Semi-professional     & 44.1      & Semi-professional quality    & yes \\
\midrule
Radio Drama               & 44.1      & Radio drama                  & no  \\
Mandarin Podcast 2        & 44.1      & Podcast in Mandarin          & no  \\
Digitized Cassette        & 32        & Digitized cassettes          & no  \\
Dipco \cite{van2019dipco}                    & 16        & Dinner party recording       & no  \\
Fleurs English  \cite{conneau2023fleurs}          & 16        & English research data        & no  \\
Librivox \cite{librivox}                  & 22.05     & Audiobook recordings         & no  \\
Old Radio                 & 22.05     & Old radio recordings         & no  \\
Phone Home                & 8         & Unscripted phone call        & no  \\
Russian Audiobook         & 44.1      & Russian audiobook            & no  \\
VSP Home Mic              & 44.1      & Home microphone audio        & no  \\
VSP Professional          & 44.1      & Professional quality         & no  \\
\bottomrule
\end{tabular}

\caption{Real audio sources with 200 samples per source for all tasks. \normalfont{ (VSP is  video sharing platform. SR is sampling rate.)}}
\label{tab:real}
\end{table}

\subsection{Setup and Rules}
The SAFE Challenge was a script-based competition. The participants did not have access to any portion of the dataset, and submissions were handled using the Hugging Face competition framework \cite{hg_comp_2025}. Participants created a private model repository on huggingface.co containing their detection model, and no restrictions were placed on what the repository could contain. This included code, model weights, and packaged environments. We made the only requirement to be a script file to read a dataset and write a submission file in a specific format. To submit a model, participants logged into the competition with their credentials to authorize the framework to access their model. Our testing framework then evaluated their model by downloading the dataset then their model, executing an evaluation script, and saving the submission file.

During the evaluation step, we sequestered the participants' models from the internet to prevent exfiltration of the competition data or their potential use of externally hosted services. To level the field, all submissions were executed on the same compute resources of an NVIDIA T4 GPU with 16GB VRAM, 32GB RAM and 8 CPUs. Additionally, all submissions were limited to 10,000 seconds, and participating teams were limited to five submissions per day. 

To further simulate real world settings, we required models to make a binary decision of either "real" or "generated" in their submission file. Therefore, the evaluation was performed at a confidence threshold chosen by each participant in each test sample. However, to facilitate further analysis, the submission file also had to contain a decision score and an average inference time for each test data point input file.

Round 1 of the competition was open for approximately 2 months. We had 12 teams participated with over 700 submissions across the three tasks. Round 2 remained active until the workshop and final presentation of results.

\subsection{Evaluation Criteria}
The main metric for the competition was balanced accuracy (BAC) defined as an average of the true positive (TPR) and the true negative rates (TNR). 
We provided limited feedback on any given submission's performance. The competition maintained both a public and a private leaderboard, and participants only had access to the public leaderboard while the competition was active. The public leaderboard showed these metrics on the entire public split as well as conditioned on every real and generated source. We anonymized the names of sources on the public leaderboard to allow for true black-box testing. The same evaluation criteria applies to each task, and the final evaluation was based on balanced accuracy over the private set.

\begin{table}[t]
\centering

\begin{tabular}{lccc}
\toprule
\textbf{Source} & \textbf{SR (kHz)} & \textbf{Voice Cloning} & \textbf{Public Split} \\
\midrule
\textbf{Cartesia} \cite{cartesia_tts_2025}       & 44.1 & Yes & yes \\
\textbf{Elevenlabs} \cite{elevenlabs_tts_2025}   & 44.1 & Yes & yes \\
Fish \cite{fish-speech-v1.4}                     & 44.1 & Yes & yes \\
Hierspeech \cite{lee2023hierspeech++}            & 48   & Yes & yes \\
Kokoro \cite{kokoro_tts_2025}                    & 24   & No  & yes \\
Parler \cite{lacombe-etal-2024-parler-tts}       & 44.1 & Yes & yes \\
Style \cite{li2023styletts}                      & 24   & Yes & yes \\
\midrule
Edge \cite{innoai_edge_tts_2025}                 & 24   & No  & no  \\
F5 \cite{chen-etal-2024-f5tts}                   & 24   & Yes & no  \\
Metavoice \cite{metavoiceio_metavoice_2025}      & 48   & Yes & no  \\
\textbf{OpenAI} \cite{openai_tts_2025}           & 24   & No  & no  \\
Seamless \cite{seamlessm4t2023}                  & 16   & No  & no  \\
Zonos \cite{steveeeeeeen_zonos_2025}             & 44.1 & Yes & no  \\
\bottomrule
\end{tabular}

\caption{Machine-generated audio with 200 samples / source for Task 1. Same models were used for Task 2 and 3. \normalfont{(Closed sources are in bold.)}}
\label{tab:generated}
\end{table}

For each task, the dataset consisted of a public and a private split. The public split was a subset of the private split. The public split consisted of 10 out of 21 real sources and 7 out of 13 TTS generator sources. The private split contained all the data. We made this choice to encourage participants to develop detectors that did not overfit to the public dataset.

\section{Dataset Description}
\subsection{Sourcing Real Audio}
The real audio for the SAFE Challenge contained $21$ diverse sources of audio. All  audio was kept in its original format (codec and sampling rate). The real audio portion of the dataset contained 200 samples per source for a total of 4,200 audio sources and 18.25 hours of data. The average length per source was .86 hours, and the average clip length was 15.64 seconds. This data source summary, along with additional metrics and descriptions, are listed in Table~\ref{tab:real}. Almost half (10 out 21 sources) were included in the public split.

The sources spanned many common ways audio is consumed, including face-to-face speech, phone calls, radio, podcasts, and audio from video sharing platforms. They also contained speech from multiple languages, including Arabic, English, German, Japanese, Mandarin, and Russian. Audio quality varied from high quality podcasts and radio dramas to telephone audio and digitized cassettes. The diversity of sourcing and audio quality forced participants to build detectors that are robust to a wide distribution of real data. 

\begin{table}[t]
\centering
\begin{tabular}{ll}

\toprule
\textbf{Augmentation} & \textbf{Description} \\
\midrule
AAC 16k              & AAC compr. with a 16kbps \\
MP3-AAC 16k          & Chained MP3 and AAC compr., 16kbps \\
Opus 16k             & Opus compr. with 16kbps \\
Resample Up          & Resample up to 48 kHz \\
Time Stretch         & Sped up while maintaining pitch \\
Encodec              & Neural codec \\
MP3-AAC-MP3 16k      & Chained MP3, AAC, MP3 compr., 16kbps \\
Phone Audio          & G.722 compr., 16kbps, sampled at 8 kHz \\
Semanticodec         & Neural codec \\
Unaugmented          & Original generated audio \\
Focalcodec           & Neural codec \\
MP3 VBR              & MP3 compr. with variable bit rate \\
Pitch Shift          & Shift the pitch up and down \\
Snac                 & Neural codec \\
Vorbis 16k           & Vorbis compr. with 16kbps \\
MP3 16k              & MP3 compr. with 16kbps \\
Noise                & Add Gaussian noise \\
Resample Down        & Resample down to 16 kHz \\
Speech Filter        & Band-pass filter from 50–7000 Hz \\
\bottomrule
\end{tabular}
\caption{Post-processing used on generated audio in Task 2.}
\label{tab:aug_desdr}
\end{table}

\begin{table}[t]
\centering
\tabfontsize
\begin{tabular}{ll}

\toprule
\textbf{Laundering Technique} & \textbf{Description} \\
\midrule
Car                  & Real car background noise \\
Reverb               & Added reverberation \\
Over Air             & Played back over the air\\
Car-Reverb Over Air  & Added car noise, applied reverb, and recorded over air \\
\bottomrule
\end{tabular}
\caption{Laundering applied to generated audio in Task 3. }
\label{tab:laundering_descr}
\end{table}


\subsection{Machine Generated Audio}
{\bf Generators:} The machine generated audio for the SAFE Challenge was built from $13$ high-performing text-to-speech models. See Table \ref{tab:generated} for more details. Only 7 out of 13 models were included in the public split. The models were a mix of proprietary (Cartesia, Edge, ElevenLabs, and OpenAI) and open source. We created 200 samples per generated source for a total of 2,600 audio sources and 8.8 hours of data. The average length per source was .68 hours, and the average clip length was 12.24 seconds. The the sample rate of clips and whether or not the model supports voice cloning is displayed in Table \ref{tab:generated}. Where available, we saved the audio as uncompressed WAVE files at the native sampling rate of the model.

{\bf Post-processing:} For Task $2$ of the SAFE Challenge, we applied $18$ different augmentations, mostly consisting of common compression schemes, to the generated dataset, while the real data remains the same. We chose these augmentations to mimic processes that could reasonably have been applied to audio on the internet. For each model, we randomly sampled $20$ clips from those generated for Task $1$, and we applied $18$ different compressions to these $20$ clips for a total of $380$ samples per model, including the unaugmented clips. In total, the augmented data contained 16.8 hours of audio with an average clip length of 12.25 seconds. The augmentations included well-known compression algorithms such as AAC, MP3, Opus, and Vorbis~\cite{valin2012rfc, moffitt2001ogg}. The augmentations also included resampling up and down, neural codecs, a speech filter, pitch shift, Gaussian noise, speeding up the audio, and a compression meant to mimic phone audio~\cite{liu2024semanticodec, della2025focalcodec, siuzdak2024snac, defossez2022highfi}. Table \ref{tab:aug_desdr} contains a list of augmentations applied to the audio files, which were applied to the public and private splits.

{\bf Laundering:} In Task $3$, we applied four different laundering techniques shown in  Table~\ref{tab:laundering_descr} to the generated audio,
while keeping the real audio the same. 
The first technique added real-life car background noise to the generated audio clips. The second technique played the audio clips over the air and recorded them. The third technique added reverberation to the generated audio. The final technique combined the previous techniques by first adding the car background noise, adding reverberation, then playing that back over the air while recording. For each source, we created 50 samples, totaling 200 samples across the 4 augmentations. Across all 13 sources, 2,600 total samples were generated, corresponding to 8.8 hours of audio. The average clip length was 12.2 seconds. 



\begin{table}[t]
\centering
\tabfontsize
\begin{tabular}{lll}
\toprule
\textbf{Code} & \textbf{Name} & \textbf{Institution} \\
\midrule
ISP & ISPL & Politecnico di Milano \\
VIP & Viper-Purdue & Purdue University \\
JAI & JAIST-HIS & Japan Advanced Institute of Science and Tech, \\
ANO & Anon\_Peking & Beijing/University of Chinese Academy of Sciences \\
DMF & DMF &  Hangzhou Dianzi University\\
\bottomrule
\end{tabular}
\caption{Details on the top five team}
\label{tab:teams}

\end{table}
\begin{table}[t]
\centering
\begin{tabular}{l|ccc|ccc|ccc}
\toprule
\textbf{Team} &
\multicolumn{3}{c|}{\textbf{Task 1}} &
\multicolumn{3}{c|}{\textbf{Task 2}} &
\multicolumn{3}{c}{\textbf{Task 3}} \\
\cmidrule(lr){2-4} \cmidrule(lr){5-7} \cmidrule(lr){8-10}
& \textbf{TPR} & \textbf{TNR} & \textbf{BAC}
& \textbf{TPR} & \textbf{TNR} & \textbf{BAC}
& \textbf{TPR} & \textbf{TNR} & \textbf{BAC} \\
\midrule
ISP          & 0.79 & \bf{0.95} & \bf{0.87} & 0.67 & \bf{0.93} & \bf{0.80} & 0.39 & \bf{0.92} & \bf{0.66} \\
VIP  & 0.74 & 0.80 & 0.77 & 0.44 & 0.90 & 0.67 & 0.60 & 0.50 & 0.55 \\
JAI     & 0.46 & 0.90 & 0.68 & 0.67 & 0.77 & 0.72 & 0.55 & 0.76 & 0.65 \\
ANO   & 0.77 & 0.71 & 0.74 & 0.84 & 0.71 & 0.78 & 0.41 & 0.72 & 0.57 \\
DMF           & \bf{0.86} & 0.49 & 0.67 & \bf{0.92} & 0.23 & 0.58 & \bf{0.81} & 0.21 & 0.51 \\
\bottomrule
\end{tabular}
\caption{True Positive Rate (TPR), True Negative Rate (TNR), Balanced Accuracy (BAC) for detection of generated audio.}
\label{tab:merged_task_results}

\end{table}
%

\section{Round 1 Results}
An overview of the Round 1 results, focusing on the binary predictions produced by participants' models is presented below. 

Table \ref{tab:merged_task_results} shows Round 1 competition results for the top five performing teams\footnote{ranked by balanced accuracy on the private split for the Task 1 at the end of Round 1}. Table~\ref{tab:teams} provides the team names and their institution. Generally, we observed an expected trend of balanced accuracy decreasing as tasks became more difficult. From Table~\ref{tab:merged_task_results}, we observe that detecting unprocessed generated audio (Task 1) followed the expected outcome of being the easiest task, with top performer (ISP) achieving a balanced accuracy (BAC) of~$.87$, True Positve Rate (TPR) of $.79$ and True Negative Rate (TNR) of $.95$. However, when we applied common compression codecs and resampling, the TPR dropped to $.67$ while maintaining a TNR of $.93$. The laundering  methods had the most drastic effect in detector performance with TPR dropping to $.39$. Recall, that for Tasks 2 and 3, the real data remained unchanged from Task 1, only the generated data was modified. Figure~\ref{fig:teaser} shows the same trend of decreasing performance over tasks with full 
ROC 
curves per team (each dot corresponds to the team's chosen decision threshold). 
The full ROC curves provide insights into whether each detector was miscalibrated. For example, DMF in Task 1 operated at a highly unbalanced threshold corresponding to $.67$ BAC. Reducing FPR would improve BAC to $.7$.


\subsection{Task Specific Discussion}

{\bf Task 1 Results.} The performance of the top 5 teams for Task 1 is displayed in Table~\ref{tab:task1_audio_accuracy_gen}, which shows BAC conditioned on the specific generator model. This was computed by averaging TPR conditioned on the generator and TNR on all real data from Table \ref{tab:merged_task_results}. Additionally, Table~\ref{tab:task1_audio_accuracy_prist} shows BAC conditioned on the source of real data computed in an analogous manner. 
Top submissions performed surprisingly well across a large collection of recent generators and diverse real sources. Performing this well is impressive considering that none of the details of the competition were unknown apriori. Even during the competition the specifics of the sources and models were not shared. The following tables show Task 1 performance.


\begin{table}[t]
\centering

\begin{tabular}{lrrrrrc}
\toprule
 \textbf{Model} & \textbf{ISP} & \textbf{VIP} & \textbf{JAI} & \textbf{ANO} & \textbf{DMF} & \textbf{Public Split} \\
\midrule
Elevenlabs  & \textbf{0.97} & 0.88 & 0.82 & 0.58 & 0.71 & yes \\
Fish        & \textbf{0.94} & 0.79 & 0.81 & 0.91 & 0.62 & yes \\
Hierspeech  & 0.76 & 0.62 & 0.62 & \textbf{0.86} & 0.63 & yes \\
Kokoro      & \textbf{0.98} & 0.90 & 0.84 & 0.80 & 0.73 & yes \\
Parler      & \textbf{0.97} & 0.74 & 0.80 & 0.48 & 0.66 & yes \\
Seamless    & 0.93 & 0.90 & 0.86 & \textbf{0.94} & 0.75 & yes \\
Style       & \textbf{0.82} & 0.71 & 0.46 & 0.46 & 0.75 & yes \\
\midrule
Cartesia    & \textbf{0.91} & 0.84 & 0.81 & 0.67 & 0.72 & no \\
Edge        & 0.68 & 0.83 & \textbf{0.85} & 0.73 & 0.73 & no \\
F5          & \textbf{0.90} & 0.67 & 0.72 & 0.50 & 0.56 & no \\
Metavoice   & \textbf{0.88} & 0.80 & 0.76 & 0.56 & 0.71 & no \\
OpenAI      & 0.86 & 0.85 & 0.75 & \textbf{0.92} & 0.72 & no \\
Zonos       & \textbf{0.71} & 0.48 & 0.56 & 0.45 & 0.52 & no \\
\midrule
Average Bal. Acc & \textbf{0.87} & 0.77 & 0.74 & 0.68 & 0.67 &  \\
\bottomrule
\end{tabular}
\caption{Balanced accuracy conditioned on the generation model in Task 1. TNR from Table \ref{tab:merged_task_results} is used in the calculation.}

\label{tab:task1_audio_accuracy_gen}

\end{table}

Generated audio from newer models (such as Zonos \cite{steveeeeeeen_zonos_2025} and Edge \cite{innoai_edge_tts_2025} with BAC $.71$ and $.68$ by ISP team) were the hardest to detect, while popular models that have been available for some time (such as ElevenLabs and Seamless with ISP BAC of $.97$ and $.93$) were detected more easily. Interestingly, older but more obscure models (such as Hierspeech with BAC $.76$) were also difficult.

For real audio, some of the rare non-English audio, such as the Arabic Speech Corpus \cite{halabi2016arabic} (ISP had BAC of $.62$), Japanese shortwave radio, and Russian audio books (VIP had BAC of $.65$ and $.56$), formed the hardest challenges to detect, potentially due to the dominance of other large resource languages in training. Poor recording quality in older audio, such as the Phone Home source (VIP had BAC of $.69$), also gave the detectors more difficulty.

In both tables, the upper rows represent models and sources from the public split, while the lower rows correspond to those exclusive to the private split. Because participants could iteratively improve their algorithms through repeated submissions, we would expect performance on models from the public split to be higher. We saw some evidence of this in Table~\ref{tab:task1_audio_accuracy_gen} but the difference was not drastic suggesting that either the algorithms do generalize or the participants had private split models in their training sets. On the reals, we did not see any clear performance difference between private and public splits.

\begin{table}[t]
\centering

\begin{tabular}{lrrrrrc}
\toprule
\textbf{Source} & \textbf{ISP} & \textbf{VIP} & \textbf{JAI} & \textbf{ANO} & \textbf{DMF} & \textbf{Public Split} \\
\midrule
Mandarin Podcast 1        & \textbf{0.90} & 0.87 & 0.89 & 0.68 & 0.61 & yes \\
Fleurs German             & \textbf{0.90} & 0.86 & 0.69 & 0.72 & 0.83 & yes \\
VSP Semi-professional      & \textbf{0.90} & 0.80 & 0.89 & 0.71 & 0.65 & yes \\
VSP Phone Call            & \textbf{0.86} & 0.78 & 0.84 & 0.73 & 0.64 & yes \\
VSP Documentary           & \textbf{0.87} & 0.84 & 0.81 & 0.66 & 0.69 & yes \\
Arabic Speech Corpus      & 0.62 & 0.38 & 0.59 & \textbf{0.68} & 0.43 & yes \\
High Quality Podcasts     & \textbf{0.85} & 0.74 & 0.71 & 0.71 & 0.52 & yes \\
Japanese Shortwave        & 0.90 & 0.65 & 0.84 & 0.68 & \textbf{0.92} & yes \\
Conference                & \textbf{0.88} & 0.79 & 0.63 & 0.63 & 0.48 & yes \\
English Podcast           & \textbf{0.89} & 0.78 & 0.78 & 0.73 & 0.48 & yes \\
\midrule
Fleurs English            & \textbf{0.90} & 0.84 & 0.56 & 0.73 & 0.89 & no \\
Dipco                     & 0.88 & 0.87 & 0.54 & 0.41 & \textbf{0.90} & no \\
Digitized Cassette        & \textbf{0.90} & 0.87 & 0.89 & 0.73 & 0.87 & no \\
Librivox                  & \textbf{0.86} & 0.83 & \textbf{0.86} & 0.73 & 0.65 & no \\
Old Radio                 & \textbf{0.90} & 0.76 & 0.65 & 0.72 & 0.51 & no \\
Phone Home                & \textbf{0.89} & 0.69 & 0.69 & 0.73 & 0.83 & no \\
Russian Audiobook         & \textbf{0.89} & 0.56 & 0.48 & 0.46 & 0.53 & no \\
Mandarin Podcast 2        & 0.90 & 0.87 & 0.77 & 0.73 & \textbf{0.91} & no \\
VSP Home Mic              & \textbf{0.88} & 0.83 & \textbf{0.88} & 0.73 & 0.59 & no \\
Radio Drama               & \textbf{0.89} & 0.82 & 0.78 & 0.73 & 0.54 & no \\
VSP Professional          & \textbf{0.89} & 0.76 & 0.79 & 0.73 & 0.67 & no \\
\midrule
Average Bal. Acc.         & \textbf{0.87} & 0.77 & 0.74 & 0.68 & 0.67 &  \\
\bottomrule
\end{tabular}
\caption{Balanced accuracy conditioned on the real source in Task 1 TPR from Table \ref{tab:merged_task_results} was used in the calculation.}
\label{tab:task1_audio_accuracy_prist}

\end{table}

{\bf Task 2 Results.} 
In this task, the real audio remained unmodified, while several operations were applied to the generated audio. Table \ref{tab:augmentation_detection} shows BAC conditioned on specific operation type again with TNR computed over the real data.

The results expose several consistent failure modes across all participants. Adding Gaussian noise with a signal-to-noise ratio of 15-40dB degraded the performance of the detectors most consistently across the board (ISP BAC dropped from $.87$ in Task 1 to $.70$ while VIP dropped from $.77$ to $.45$). Speech-specific processing and codecs, such as Opus \cite{valin2012rfc}, Phone Audio, and Speech filtering, also caused a significant negative effect on detection accuracy. For example, ISP BAC dropped to $.69$, $.71$ and $.76$ respectively. Lastly, chaining multiple operations together, such as converting between MP3 and AAC codecs, degraded performance significantly as well. This shows that further research should be conducted to make detectors more robust against these post-processing operations. Furthermore, such techniques could be used 
to intentionally avoid detection by forensic systems, as our Task~3 resutls show.

Re-encoding the audio using some of the latest neural codecs (such as Encodec \cite{defossez2022highfi} and Focal codec \cite{della2025focalcodec}) did not significantly affect performance. (ISP BAC was $.92$ and $.95$). These methods' similarity to decoders used in the synthetic generation models likely caused this effect. In some cases, post-processing actually improved detector performance overall (ANO and JAI BAC improved from $.68$ and $.74$ in Task 1 to $.72$ and $.78$ in Task 2), potentially making the generated audio appear closer to other well known generators than the original pre-processed versions. 

\begin{table}[t]
\centering
\begin{tabular}{lrrrrr}
\toprule
\textbf{Augmentation} & \textbf{ISP} & \textbf{VIP} & \textbf{JAI} & \textbf{ANO} & \textbf{DMF} \\
\midrule
AAC 16k            & 0.77 & 0.72 & \textbf{0.82} & 0.63 & 0.59 \\
Encodec            & \textbf{0.92} & 0.66 & 0.85 & 0.80 & 0.58 \\
Focalcodec         & \textbf{0.95} & 0.63 & 0.83 & 0.88 & 0.57 \\
MP3\textendash AAC\textendash mp3 16k & 0.67 & 0.79 & \textbf{0.83} & 0.63 & 0.59 \\
MP3\textendash AAC 16k & 0.73 & 0.78 & \textbf{0.82} & 0.61 & 0.59 \\
MP3 16k            & \textbf{0.83} & 0.78 & 0.78 & 0.60 & 0.58 \\
MP3 VBR            & \textbf{0.84} & 0.74 & 0.76 & 0.67 & 0.58 \\
Noise              & 0.70 & 0.45 & 0.54 & 0.60 & 0.58 \\
Opus 16k           & 0.69 & 0.73 & \textbf{0.83} & 0.73 & 0.61 \\
Phone audio        & 0.71 & 0.61 & \textbf{0.79} & 0.64 & 0.56 \\
Pitch shift        & 0.85 & 0.70 & 0.77 & \textbf{0.88} & 0.60 \\
Resample down      & \textbf{0.77} & 0.63 & 0.76 & 0.68 & 0.57 \\
Resample up        & \textbf{0.82} & 0.61 & 0.76 & 0.68 & 0.57 \\
Semanticodec       & 0.83 & 0.67 & 0.85 & \textbf{0.89} & 0.58 \\
Snac               & 0.78 & 0.65 & 0.81 & \textbf{0.87} & 0.60 \\
Speech filter      & \textbf{0.76} & 0.61 & 0.67 & 0.66 & 0.45 \\
Time stretch       & 0.85 & 0.69 & 0.78 & \textbf{0.89} & 0.61 \\
Vorbis 16k         & \textbf{0.86} & 0.67 & 0.75 & 0.67 & 0.57 \\
\midrule
Average Bal. Acc.  & \textbf{0.80} & 0.67 & 0.78 & 0.72 & 0.58 \\
\bottomrule
\end{tabular}

\caption{Balanced accuracy conditioned on the augmentation type applied to generated data in Task 2. 
}
\label{tab:augmentation_detection}

\end{table}

{\bf Task 3 Results.} 
This task presented the most difficult challenge. Table \ref{tab:playback_reverb_conditions_colored} shows BAC conditioned on the laundering method applied only to the generated samples. Here, we used our existing benchmark detectors to select a set of operations that maximally decreased the detection rate. These include adding real-life car background noise, applying reverb, and recording sound played over the air. While all laundering techniques were effective on their own, the combination of all three was the most difficult to detect. Even on its own, recording synthetic audio being played over the air resulted in significant degradation. ISP BAC dropped from $.87$ on unprocessed samples in Task 1 to $.62$ while VIP dropped from $.77$ to $.50$ This effect likely occurred due to reintroducing the artifacts of the complete recording pipeline, including the computer analog-to-digital converter, speaker amplifier, acoustic environment, microphone amplifier, and digital-to-analog, etc. While this laundering technique is not scalable, it proved highly effective. These observations point to a potential and easily exploitable vulnerability in the detection of synthetic audio using current methods.

\section{Conclusion}
In this paper, we provided an overview of the Synthetic Audio Forensics Evaluation (SAFE) Challenge that tested the robustness of synthetic audio detectors against a diversity of real sources, unknown audio generation models, and benign and malicious audio processing operations. We ran and detailed a unique audio forensic competition where participants were not given training data, 
had limited knowledge of how data was created, and were provided with limited performance feedback. Under these difficult conditions, several submissions achieved high accuracy with limited degradation under benign audio processing. However, targeted laundering still resulted in significant reduction in detection performance.

\begin{table}[t]
\centering
\begin{tabular}{lrrrrr}
\toprule
\textbf{Laundering Technique} & \textbf{ISP} & \textbf{VIP} & \textbf{JAI} & \textbf{ANO} & \textbf{DMF} \\
\midrule
Car                & \textbf{0.67} & 0.50 & 0.53 & 0.62 & 0.57 \\
Played             & 0.62          & 0.54 & 0.54 & \textbf{0.69} & 0.54 \\
Reverb             & \textbf{0.75} & 0.46 & 0.70 & 0.59 & 0.58 \\
Played + Reverb + Car  & 0.58          & 0.69 & 0.49 & \textbf{0.72} & 0.35 \\
\midrule
Average Bal. Acc.  & \textbf{0.66} & 0.55 & 0.57 & 0.65 & 0.51 \\
\bottomrule
\end{tabular}

\caption{Balanced accuracy conditioned on the laundering type applied to generated data in Task 3. 
}

\label{tab:playback_reverb_conditions_colored}

\end{table}

\section{Acknowledgments}
This material is based on research sponsored by ULRI DSRI. 

\bibliographystyle{unsrt}  
\bibliography{references_short}

\appendix
\section{Round 2 Discussion}
\begin{figure}[htb]
\centering
    \fbox{\includegraphics[width=.75 \textwidth]
    {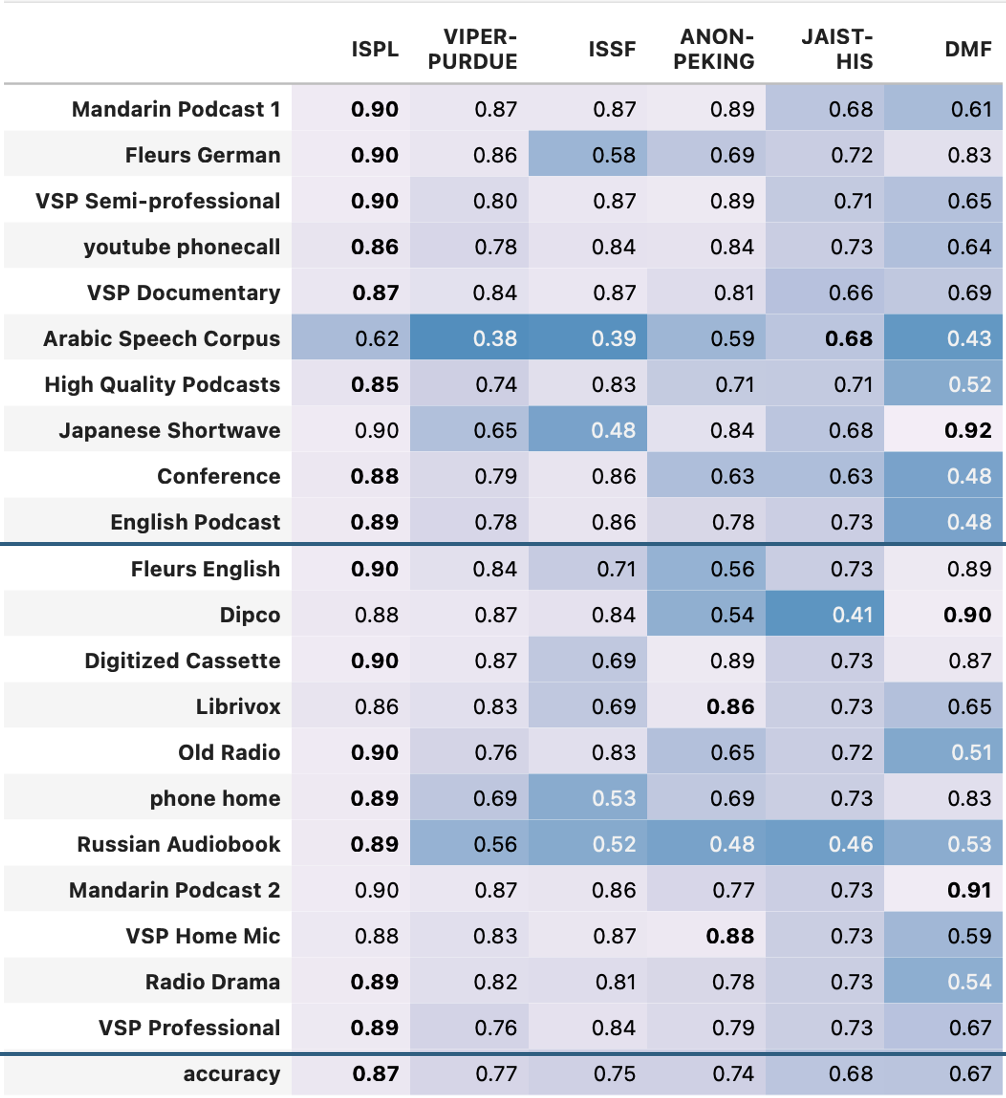}}
    \caption{Task 1 TNR results conditioned on real source in Round 2.}
    \label{fig:task1_real_round2}
\end{figure}

\begin{figure}[htb]
\centering
    \fbox{\includegraphics[width=.75 \textwidth]
    {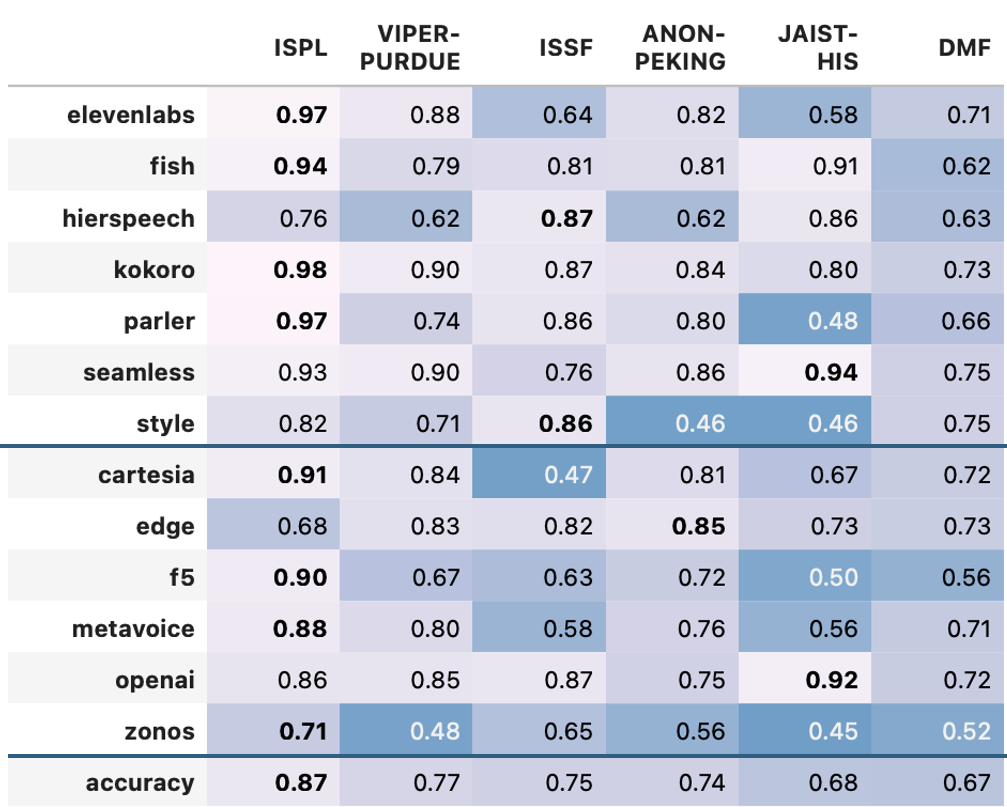}}
    \caption{Task 1 TPR results conditioned on generated source in Round 2.}
    \label{fig:task1_generated_round2}
\end{figure}

\begin{figure}[htb]
\centering
    \fbox{\includegraphics[width=.75 \textwidth]
    {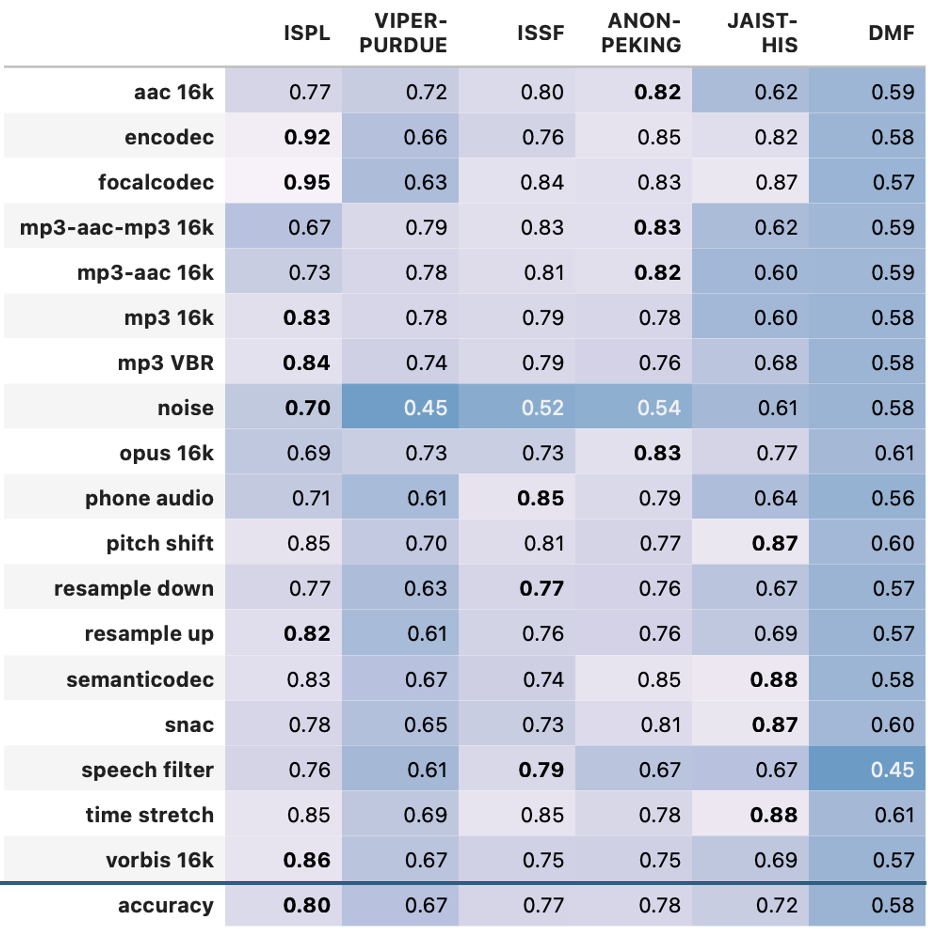}}
    \caption{Task 2 balanced accuracy conditioned on augmentation in Round 2.}
    \label{fig:task2_round2}
\end{figure}

\begin{figure}[htb]
\centering
    \fbox{\includegraphics[width=.75 \textwidth]
    {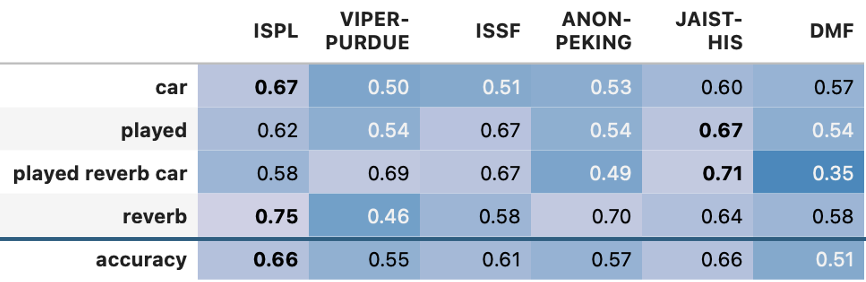}}
    \caption{Task 3 balanced accuracy conditioned on laundering technique in Round 2.}
    \label{fig:task3_round2}
\end{figure}

The final results from Round 2 follow the same trends as Round $1$, with balanced accuracy decreasing across subsequent tasks. The teams remain the same as Round $1$ with the addition of ISSF from University of Michigan. The top performer for Tasks 1-3 was still ISPL with a balanced accuracies of $.87$, $.80$ and $.66$. The second best performing models in Tasks 2 and 3 differed from ISPL by only a few percentage points. The updated scores for Task 1 on the real and generated data from Round 2 are located in Figures \ref{fig:task1_real_round2} and \ref{fig:task1_generated_round2}, respectively.
For Task 2, the updated scores for all augmentation techniques from Round 2 are located in Figure \ref{fig:task2_round2}.
Lastly, The updated scores for Task 3 for all laundering techniques from Round 2 are located in Figure \ref{fig:task3_round2}. \par
\begin{figure}
\centering
    \fbox{\includegraphics[width=.95 \textwidth]
    {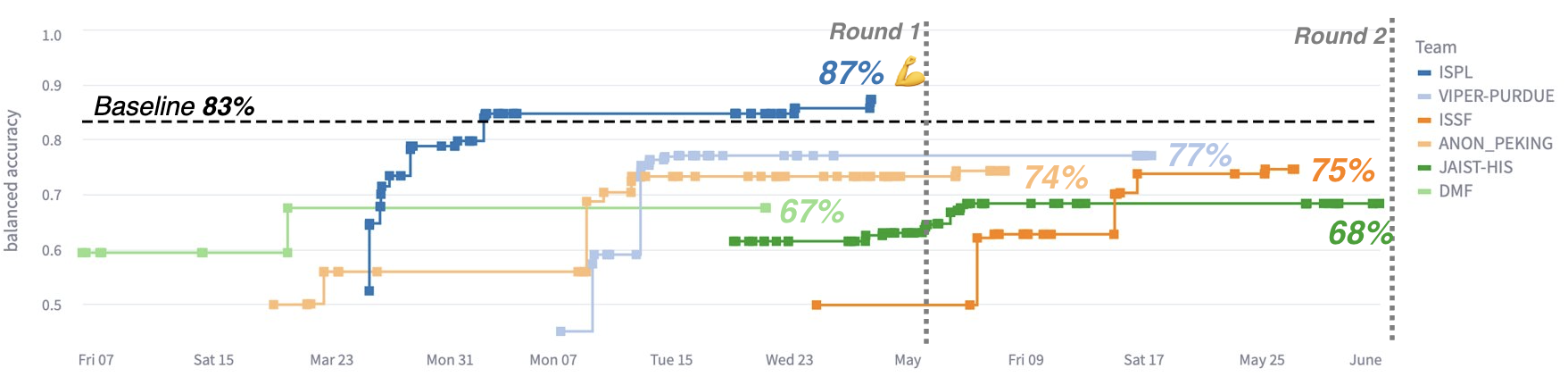}}
    \caption{Balanced accuracy over time in Task 1.}
    \label{fig:performance_timeline}
\end{figure}

\begin{figure}
\centering
    \fbox{\includegraphics[width=.75 \textwidth]
    {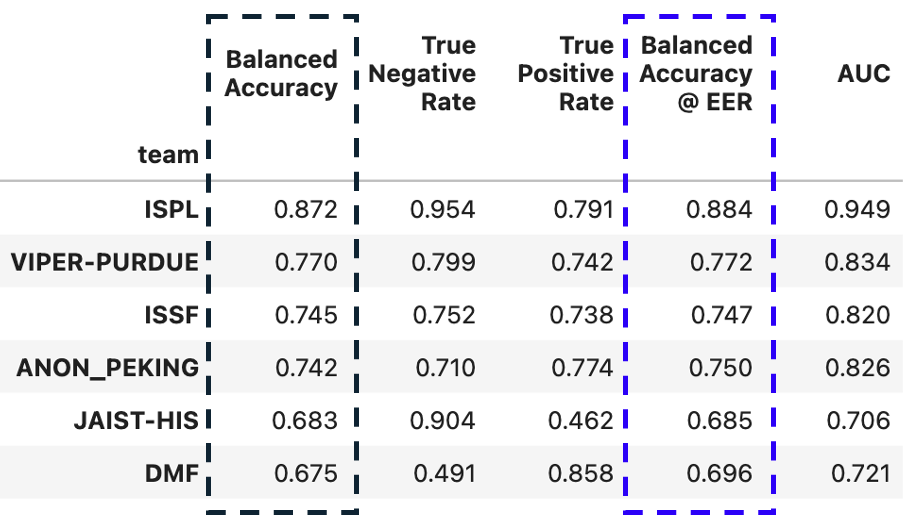}}
    \caption{Balanced accuracy at the equal error rate for Task 1.}
    \label{fig:calibration_table}
\end{figure}

All teams exhibited significant improvements in their balanced accuracy over time, as depicted in Figure \ref{fig:performance_timeline}. The highest performing team, ISPL, improved from a balanced accuracy of $.52$ to $.87$ between their initial and final submissions. Other teams had similar levels of improvement between their first and last submissions and exhibited improvements deep into Round 2. In every case, the highest initial balanced accuracy was below $.65$. This highlights one weakness of state of the art models currently: difficulty generalizing. For real world applications, there may not be ground truth available to use to retrain or recalibrate the models.
There was also some miscalibration among the algorithms with some favoring lower false positive rates over higher true positive rates, as seen in Figure \ref{fig:calibration_table}. However, DMF argued that in real applications not detecting generated audio is usually more costly than a false alarm.

\subsection{Training Data Used By Teams}

The two main factors that could affect team performance are the training data and model architectures used. The SAFE dataset was curated specifically to be varied and encourage the participants to build models that generalize well. The participating teams tended to use different training data from each other, however, ASVspoof2019 \cite{todisco2019asvspoof} was used by three different teams. Two teams used datasets curated in-house with one team relying only on that dataset. ISPL trained on the most datasets, providing anecdotal evidence that a wider set of training data boosts performance. ISSF used a systematic iterative approach to their dataset generation, starting with a single source ASVspoof2019 and increasing the number of sources they trained on while tracking performance. They chose the added datasets to specifically augment identified weaknesses in the previous version of their training dataset. The training datasets used by the teams who presented at IH\&MMSEC 2025 are in Table \ref{tab:training_datasets}.

\begin{table}[htb]
\centering
\renewcommand{\arraystretch}{1.2}
\begin{tabular}{|l|c|c|c|c|c|}
\hline
\textbf{Dataset} & \textbf{ANO} & \textbf{DMF} & \textbf{JAI} & \textbf{ISPL} & \textbf{ISSF} \\ \hline

ASVspoof2021 \cite{liu2023asvspoof} & X &  &  &  &  \\ \hline
ADD 2023 Track 1.2 Test R2 \cite{yi2023add} & X &  &  &  &  \\ \hline
CtrSVDD 2024 \cite{zhang2024svdd} & X &  &  &  &  \\ \hline
Emilia dataset \cite{he2024emilia} & X &  &  &  &  \\ \hline
CosyVoice2 \cite{du2024cosyvoice} & X &  &  &  &  \\ \hline
ASVspoof2019 \cite{todisco2019asvspoof} &  & X &  & X & X \\ \hline
Codecfake \cite{xie2025codecfake} &  & X &  &  & X \\ \hline
CFAD \cite{ma2024cfad} &  & X &  &  &  \\ \hline
WaveFake \cite{frank2021wavefake} &  & X &  &  &  \\ \hline
In-the-Wild \cite{muller2022does} &  & X &  & X &  \\ \hline
JMAD dataset \cite{mawalim2025multilingual} &  &  & X &  &  \\ \hline
Fake-or-Real \cite{foR_dataset_2019} &  &  &  & X &  \\ \hline
MLAAD \cite{muller2024mlaad} &  &  &  & X & X \\ \hline
DiffSSD \cite{bhagtani2025diffssd} &  &  &  & X &  \\ \hline
LibriSpeech \cite{panayotov2015librispeech} &  &  &  & X &  \\ \hline
LJSpeech \cite{xu2020lrspeech} &  &  &  & X &  \\ \hline
VCTK \cite{awsaf49_vctk_sr16k_dataset} &  &  &  & X &  \\ \hline
Mozilla CommonVoice \cite{ardila2019common} &  &  &  & X &  \\ \hline
SpoofCeleb \cite{jung2025spoofceleb} &  &  &  &  & X \\ \hline
M-AILABS \cite{solak2019m} &  &  &  &  & X \\ \hline
Famous Figures (in-house dataset) &  &  &  &  & X \\ \hline

\end{tabular}
\caption{Training datasets used by each team.}
\label{tab:training_datasets}
\end{table}
\subsection{Training Data Used By Participants}
There was some overlap between some of the performers in their choice of model architecture.
Three of the five performers who presented at IH\&MMSEC, ANO, JAI and ISSF used Aasist \cite{jung2022aasist} as the base model for their detectors which is a GAN-based speech detection system. ANO combined Aasist with a sample weight learning module to help combat issues with distribution shift. JAI combined RawNet2 \cite{tak2021end} with Aasist and self-supervised-learning techniques. ISSF used a self-supervised-learning front-end for feature extraction and an Aasist backend for classification. DMF, like JAI, used RawNet2 as part of their detector pipeline. DMF tested various combinations of label noise learning (LNL), inter-sample distillation (ISD) and SSI (self-supervised initialization) as the input into the RawNet2 framework. 
ISPL, the highest performer, used a mixture-of-experts approach (MOE) for their detector. Specifically, they used a mixture of implicitly localized experts (MILE) where each expert has a different model architecture, but the training data is the same across experts. ISPL settled on using three experts using LCNN + MelSpec, Resnet + MelSpec and ResNet + LogSpec. 

\end{document}